\documentclass[pra]{revtex4}
 \usepackage{amssymb} \usepackage{graphicx}

\newtheorem{theorem}{Theorem}

\begin{document}
 \title{Twistor spinors and extended conformal superalgebras}

\author{\"Umit Ertem}
 \email{umitertemm@gmail.com}
\address{Astronomer, Diyanet \.{I}\c{s}leri Ba\c{s}kanl{\i}\u{g}{\i}, \"{U}niversiteler Mah.,\\ Dumlup{\i}nar Bul. No:147/H 06800 \c{C}ankaya, Ankara, Turkey\\}

\begin{abstract}

We show that the first-order symmetry operators of twistor spinors can be constructed from conformal Killing-Yano forms in conformally-flat backgrounds. We express the conditions on conformal Killing-Yano forms to obtain mutually commuting symmetry operators of twistor spinors. Conformal superalgebras which consist of conformal Killing vectors and twistor spinors and play important roles in supersymmetric field theories in conformal backgrounds are extended to more general superalgebras by using the graded Lie algebra structure of conformal Killing-Yano forms and the symmetry operators of twistor spinors. The even part of the extended conformal superalgebra corresponds to conformal Killing-Yano forms and the odd part consists of twistor spinors.

\end{abstract}

\maketitle

\section{Introduction}

Twistor spinors are solutions of a special type of spinor equation that first appeared in the context of general relativity \cite{Penrose Rindler, Lichnerowicz1, Lichnerowicz2}. Since the twistor equation is conformally invariant, twistor spinors are related to the conformal symmetries of the background \cite{Habermann1, Baum Friedrich Grunewald Kath, Baum}. Conformal Killing vector fields of a background can be constructed from the Dirac currents of twistor spinors. Special types of twistor spinors are related to the supersymmetry parameters of different supergravity theories in various dimensions \cite{de Medeiros}. Because of the conformal covariance of the twistor equation, some rigid conformal supersymmetric field theories can also be defined on the backgrounds they exist. Classification of the backgrounds that admit twistor spinors is studied in Riemannian and Lorentzian signatures \cite{Baum Leitner}.

Supergravity backgrounds are characterized by some Lie superalgebras defined on them which are called Killing superalgebras \cite{OFarrill Meessen Philip, OFarrill HackettJones Moutsopoulos}. They consist of Killing vectors as the even part and Killing spinors as the odd part of the superalgebra \cite{Klinker}. This superalgebra construction can be extended to the case of conformal Killing vectors and twistor spinors \cite{Habermann2}. The defined conformal superalgebra does not correspond to a Lie superalgebra in general and extra $R$-symmetries can be included in it to obtain a Lie superalgebra structure \cite{de Medeiros Hollands, Lischewski}. Conformal superalgebras are constructed from the Dirac currents of twistor spinors which correspond to conformal Killing vectors, the Lie algebra of conformal Killing vectors and the Lie derivative of twistor spinors with respect to conformal Killing vectors. They are related to rigid supersymmetric field theories in conformal supergravity backgrounds since the supersymmetry parameters of those theories correspond to twistor spinors \cite{Cassani Klare Martelli Tomasiello Zaffaroni}.

In this paper, we construct the extensions of conformal superalgebras to include conformal Killing-Yano (CKY) forms which are antisymmetric generalizations of conformal Killing vector fields to higher-degree forms. Besides the Dirac currents of spinors, higher-degree $p$-form Dirac currents of them can also be defined from the Fierz identity that gives way to write the tensor product of a spinor and its dual as a sum of different degree differential forms. It is known that the $p$-form Dirac currents of twistor spinors correspond to CKY forms \cite{Acik Ertem}. We show that the Lie derivative of twistor spinors with respect to conformal Killing vectors can be generalized to symmetry operators in terms of CKY forms. The constructed operators are symmetry operators of twistor spinors in constant curvature backgrounds for all CKY forms and in Einstein manifolds for normal CKY forms. This is also true for all CKY forms in conformally-flat backgrounds. These results are considered as a first step for a full understanding of the symmetries of the twistor operator where conformal invariance must play an essential role. In the existence of mutually commuting symmetry operators of an equation, one can obtain the solutions of it by the method of separation of variables. We calculate the commutator of symmetry operators of twistor spinors and find the conditions to obtain mutually commuting symmetry operators. On the other hand, CKY forms in constant curvature backgrounds and normal CKY forms in Einstein manifolds satisfy a graded Lie algebra structure as proven in \cite{Ertem1}. So, by using the generalized symmetry operators of twistor spinors, graded Lie algebra of CKY forms and the $p$-form Dirac currents of twistor spinors, we obtain the extended conformal superalgebras of CKY forms and twistor spinors in conformally-flat manifolds. These extended superalgebras can be related to different rigid supersymmetric field theories in conformal backgrounds. Since Killing-Yano (KY) forms consist of a subset of CKY forms and Killing spinors consist of a subset of twistor spinors, the extended conformal superalgebras reduce to extended Killing superalgebras for the case of KY forms and Killing spinors \cite{Ertem2}.

The paper is organized as follows. We show the construction of symmetry operators of twistor spinors in terms of CKY forms in constant curvature backgrounds, Einstein manifolds and conformally-flat backgrounds in Section 2. We also show the conformal invariance of the constructed symmetry operators and calculate the commutator of symmetry operators to find the conditions for obtaining mutually commuting symmetry operators in two subsections. In Section 3, we construct the extended conformal superalgebras and discuss the reduction to extended Killing superalgebras. Section 4 concludes the paper.

\section{Symmetry Operators of Twistor Spinors}

Spin geometry deals with geometric structures on manifolds that admit spin structure. The main first-order differential operators in spin geometry are Dirac operator and twistor operator. Dirac operator is defined in terms of the frame basis ${X_a}$, co-frame basis ${e^a}$ which is defined as $e^a(X_b)=\delta^a_b$ and the spinor covariant derivative $\nabla_{X_a}$ as follows
\begin{equation}
\displaystyle{\not}D=e^a.\nabla_{X_a}
\end{equation}
where $.$ denotes the Clifford product. Here and in the following formulas we use Einstein summation convention. We can raise and lower the indices of $X_a$ and $e^a$ by using the components of the metric $g_{ab}$ and the inverse metric $g^{ab}$ as $X^a=g^{ab}X_b$ and $e_a=g_{ab}e^b$. Co-frame basis $e^a$ corresponds to the basis of the Clifford bundle and we have the identity
\[
e^a.e^b+e^b.e^a=2g^{ab}.
\]
A spinor $\psi$ is called a harmonic spinor if it is in the kernel of the Dirac operator
\begin{equation}
\displaystyle{\not}D\psi=0
\end{equation}
and this equation is called the massless Dirac equation. Twistor operator is defined with respect to any vector field $X$ and its metric dual $\widetilde{X}$ in $n$ dimensions as
\begin{equation}
{\cal{P}}_X:=\nabla_X-\frac{1}{n}\widetilde{X}.\displaystyle{\not}D.
\end{equation}
Spinors that are in the kernel of the twistor operator, namely that satisfy the following twistor equation
\begin{equation}
\nabla_X\psi=\frac{1}{n}\widetilde{X}.\displaystyle{\not}D\psi
\end{equation}
are called twistor spinors.

An operator that takes a solution of a differential equation and gives another solution is called a symmetry operator of that equation. Symmetry operators of massless Dirac equation can be constructed from the Lie derivative of spinor fields with respect to conformal Killing vectors. In a background with metric $g$, a conformal Killing vector $K$ preserves the metric up to conformal rescalings
\begin{equation}
{\cal{L}}_Kg=2\lambda g
\end{equation}
where ${\cal{L}}_K$ is the Lie derivative with respect to $K$ and $\lambda$ is a function. On the other hand, the Lie derivative on spinor fields is defined as follows
\begin{equation}
{\cal{L}}_K\psi=\nabla_K\psi+\frac{1}{4}d\widetilde{K}.\psi
\end{equation}
where $d$ is the exterior derivative operator. The following operator is a symmetry operator for massless Dirac equation
\begin{equation}
{\cal{L}}_K+\frac{1}{2}(n-1)\lambda
\end{equation}
and hence preserves the space of harmonic spinors \cite{Benn Tucker}. Moreover, CKY forms which are antisymmetric generalizations of conformal Killing vector fields to higher degree forms can also be used in the construction of symmetry operators of massless Dirac equation. A $p$-form $\omega$ is a CKY $p$-form if it satisfies the following equation
\begin{equation}
\nabla_X\omega=\frac{1}{p+1}i_Xd\omega-\frac{1}{n-p+1}\widetilde{X}\wedge\delta\omega
\end{equation}
where $i_X$ is the interior derivative or contraction operator with respect to $X$ and $\delta$ is the co-derivative operator. So, the most general first-order symmetry operator of the massless Dirac equation is written in terms of CKY $p$-forms as \cite{Benn Charlton}
\begin{equation}
L_{\omega}=(i_{X^a}\omega).\nabla_{X_a}+\frac{p}{2(p+1)}d\omega-\frac{n-p}{2(n-p+1)}\delta\omega.
\end{equation}
It reduces to (7) for a CKY 1-form since the following equality is satisfied for the metric duals of conformal Killing vectors
\begin{equation}
\delta\widetilde{K}=-n\lambda.
\end{equation}

As in the massless Dirac case, the Lie derivative with respect to a conformal Killing vector is also used in the construction of symmetry operators of the twistor equation. So, the following operator preserves the solutions of the twistor equation
\begin{equation}
{\cal{L}}_K-\frac{1}{2}\lambda
\end{equation}
where $K$ and $\lambda$ are as in (5). Resembling the operator defined in (9) for the massless Dirac case, one can also search for the generalized symmetry operators for twistor spinors. One possible candidate can be defined by using CKY $p$-forms $\omega$ as follows
\begin{equation}
L_{\omega}=(i_{X^a}\omega).\nabla_{X_a}+\frac{p}{2(p+1)}d\omega+\frac{p}{2(n-p+1)}\delta\omega.
\end{equation}
It reduces to (11) for the metric dual of a CKY 1-form. For a twistor spinor $\psi$, it can be written as
\begin{eqnarray}
L_{\omega}\psi&=&(i_{X^a}\omega).\nabla_{X_a}\psi+\frac{p}{2(p+1)}d\omega.\psi+\frac{p}{2(n-p+1)}\delta\omega.\psi\nonumber\\
&=&\frac{1}{n}(i_{X_a}\omega).e^a.\displaystyle{\not}D\psi+\frac{p}{2(p+1)}d\omega.\psi+\frac{p}{2(n-p+1)}\delta\omega.\psi\nonumber\\
&=&-(-1)^p\frac{p}{n}\omega.\displaystyle{\not}D\psi+\frac{p}{2(p+1)}d\omega.\psi+\frac{p}{2(n-p+1)}\delta\omega.\psi
\end{eqnarray}
where we have used the twistor equation (4) and the definition of the Clifford product in terms of wedge product which is given for any $p$-form $\omega$ as $e^a.\omega=e^a\wedge\omega+i_{X^a}\omega$ and $\omega.e^a=e^a\wedge\eta\omega-i_{X^a}\eta\omega$. So, we have
\begin{eqnarray}
i_{X_a}\omega.e^a&=&e^a\wedge\eta i_{X_a}\omega-i_{X^a}\eta i_{X_a}\omega\nonumber\\
&=&-(-1)^pe^a\wedge i_{X_a}\omega\nonumber\\
&=&-(-1)^pp\omega.
\end{eqnarray}
Here $\eta$ is the main automorphism of the exterior algebra which acts on a $p$-form $\alpha$ as $\eta\alpha=(-1)^p\alpha$ and we have the properties $i_Xi_Y=-i_Yi_X$ and $e^a\wedge i_{X_a}\alpha=p\alpha$.

To show the symmetry operator property of (13) for twistor spinors, we need to prove that for a twistor spinor $\psi$, $L_{\omega}\psi$ also satisfies the twistor equation, namely
\begin{equation}
\nabla_{X_a}L_{\omega}\psi=\frac{1}{n}e_a.\displaystyle{\not}DL_{\omega}\psi.
\end{equation}
By using (13), the left hand side of (15) can be written as
\begin{eqnarray}
\nabla_{X_a}L_{\omega}\psi&=&-(-1)^p\frac{p}{n}\nabla_{X_a}\omega.\displaystyle{\not}D\psi-(-1)^p\frac{p}{n}\omega.\nabla_{X_a}\displaystyle{\not}D\psi+\frac{p}{2(p+1)}\nabla_{X_a}d\omega.\psi\nonumber\\
&&+\frac{p}{2(p+1)}d\omega.\nabla_{X_a}\psi+\frac{p}{2(n-p+1)}\nabla_{X_a}\delta\omega.\psi+\frac{p}{2(n-p+1)}\delta\omega.\nabla_{X_a}\psi\nonumber\\
&=&-(-1)^p\frac{p}{n}\left(\frac{1}{p+1}i_{X_a}d\omega-\frac{1}{n-p+1}e_a\wedge\delta\omega\right).\displaystyle{\not}D\psi\nonumber\\
&&-(-1)^p\frac{p}{n}\omega.\nabla_{X_a}\displaystyle{\not}D\psi+\frac{p}{2(p+1)}\nabla_{X_a}d\omega.\psi+\frac{p}{2(p+1)}\frac{1}{n}d\omega.e_a.\displaystyle{\not}D\psi\nonumber\\
&&+\frac{p}{2(n-p+1)}\nabla_{X_a}\delta\omega.\psi+\frac{p}{2(n-p+1)}\frac{1}{n}\delta\omega.e_a.\displaystyle{\not}D\psi
\end{eqnarray}
where we have used the CKY equation (8) and the twistor equation (4). The integrability condition of twistor equation is
\begin{equation}
\nabla_{X_a}\displaystyle{\not}D\psi=\frac{n}{2}K_a.\psi
\end{equation}
(see \cite{Baum Friedrich Grunewald Kath, Benn Kress}) where the 1-form $K_a$ is defined in terms of the Ricci 1-forms $P_a$ and curvature scalar ${\cal{R}}$ as
\begin{equation}
K_a=\frac{1}{n-2}\left(\frac{\cal{R}}{2(n-1)}e_a-P_a\right).
\end{equation}
By using the curvature 2-forms $R_{ab}$, the other curvature characteristics are defined as $P_a=i_{X^b}R_{ba}$ and ${\cal{R}}=i_{X^a}P_a$. So, (16) can be written in the following form
\begin{eqnarray}
\nabla_{X_a}L_{\omega}\psi&=&\bigg[-(-1)^p\frac{p}{n(p+1)}i_{X_a}d\omega+(-1)^p\frac{p}{n(n-p+1)}e_a\wedge\delta\omega\nonumber\\
&&+\frac{p}{2n(p+1)}d\omega.e_a+\frac{p}{2n(n-p+1)}\delta\omega.e_a\bigg].\displaystyle{\not}D\psi\nonumber\\
&&+\bigg[-(-1)^p\frac{p}{2}\omega.K_a+\frac{p}{2(p+1)}\nabla_{X_a}d\omega+\frac{p}{2(n-p+1)}\nabla_{X_a}\delta\omega\bigg].\psi\nonumber\\
&=&\bigg[-(-1)^p\frac{p}{2n(p+1)}i_{X_a}d\omega-(-1)^p\frac{p}{2n(p+1)}e_a\wedge d\omega\nonumber\\
&&+(-1)^p\frac{p}{2n(n-p+1)}e_a\wedge\delta\omega+(-1)^p\frac{p}{2n(n-p+1)}i_{X_a}\delta\omega\bigg].\displaystyle{\not}D\psi\nonumber\\
&&+\bigg[-(-1)^p\frac{p}{2}\omega.K_a+\frac{p}{2(p+1)}\nabla_{X_a}d\omega+\frac{p}{2(n-p+1)}\nabla_{X_a}\delta\omega\bigg].\psi\nonumber\\
\end{eqnarray}
where we have used the expansion of the Clifford product in terms of the wedge product and interior derivative.

For the right hand side of (15), we can write from (13) and the definition of the Dirac operator in (1) as
\begin{eqnarray}
\frac{1}{n}e_a.\displaystyle{\not}DL_{\omega}\psi&=&\frac{1}{n}e_a.e^b.\nabla_{X_b}\bigg(-(-1)^p\frac{p}{n}\omega.\displaystyle{\not}D\psi+\frac{p}{2(p+1)}d\omega.\psi+\frac{p}{2(n-p+1)}\delta\omega.\psi\bigg)\nonumber\\
&=&\frac{1}{n}e_a.e^b.\bigg(-(-1)^p\frac{p}{n}\nabla_{X_b}\omega.\displaystyle{\not}D\psi-(-1)^p\frac{p}{n}\omega.\nabla_{X_b}\displaystyle{\not}D\psi\nonumber\\
&&+\frac{p}{2(p+1)}\nabla_{X_b}d\omega.\psi+\frac{p}{2(p+1)}d\omega.\nabla_{X_b}\psi\nonumber\\
&&+\frac{p}{2(n-p+1)}\nabla_{X_b}\delta\omega.\psi+\frac{p}{2(n-p+1)}\delta\omega.\nabla_{X_b}\psi\bigg).
\end{eqnarray}
By using the CKY equation (8) and the integrability condition (17), one obtains
\begin{eqnarray}
\frac{1}{n}e_a.\displaystyle{\not}DL_{\omega}\psi&=&\bigg[-(-1)^p\frac{p}{n^2(p+1)}e_a.e^b.i_{X_b}d\omega+(-1)^p\frac{p}{n^2(n-p+1)}e_a.e^b.(e_b\wedge\delta\omega)\nonumber\\
&&+\frac{p}{2n^2(p+1)}e_a.e^b.d\omega.e_b+\frac{p}{2n^2(n-p+1)}e_a.e^b.\delta\omega.e_b\bigg].\displaystyle{\not}D\psi\nonumber\\
&&+\bigg[-(-1)^p\frac{p}{2n}e_.e^b.\omega.K_b+\frac{p}{2n(p+1)}e_a.e^b.\nabla_{X_b}d\omega\nonumber\\
&&+\frac{p}{2n(n-p+1)}e_a.e^b.\nabla_{X_b}\delta\omega\bigg].\psi
\end{eqnarray}
After expanding the Clifford products in terms of wedge products in the following form
\begin{eqnarray}
e_a.e^b.i_{X_b}d\omega&=&(p+1)e_a\wedge d\omega+(p+1)i_{X_a}d\omega\nonumber\\
e_a.e^b.(e_b\wedge\delta\omega)&=&(n-p+1)e_a\wedge\delta\omega+(n-p+1)i_{X_a}\delta\omega\nonumber\\
e_a.e^b.d\omega.e_b&=&-(-1)^p(n-2(p+1))e_a\wedge d\omega\nonumber\\
&&-(-1)^p(n-2(p+1))i_{X_a}d\omega\\
e_a.e^b.\delta\omega.e_b&=&-(-1)^p(n-2(p-1))e_a\wedge\delta\omega\nonumber\\
&&-(-1)^p(n-2(p-1))i_{X_a}\delta\omega\nonumber
\end{eqnarray}
(21) transforms into
\begin{eqnarray}
\frac{1}{n}e_a.\displaystyle{\not}DL_{\omega}\psi&=&\bigg[-(-1)^p\frac{p}{2n(p+1)}i_{X_a}d\omega-(-1)^p\frac{p}{2n(p+1)}e_a\wedge d\omega\nonumber\\
&&+(-1)^p\frac{p}{2n(n-p+1)}e_a\wedge\delta\omega+(-1)^p\frac{p}{2n(n-p+1)}i_{X_a}\delta\omega\bigg].\displaystyle{\not}D\psi\nonumber\\
&&+\bigg[-(-1)^p\frac{p}{2n}e_a.e^b.\omega.K_b+\frac{p}{2n(p+1)}e_a.e^b.\nabla_{X_b}d\omega\nonumber\\
&&+\frac{p}{2n(n-p+1)}e_a.e^b.\nabla_{X_b}\delta\omega\bigg].\psi
\end{eqnarray}

From (19) and (23), it can be seen that the coefficients of $\displaystyle{\not}D\psi$ are equal to each other. So, we need to show the equality of the coefficients of $\psi$ in both equations. We will consider the integrability conditions of the CKY equation for that aim. In general, the integrability conditions of the CKY equation for a $p$-form $\omega$ can be written as follows \cite{Semmelmann, Ertem1}
\begin{eqnarray}
\nabla_{X_a}d\omega&=&\frac{p+1}{p(n-p+1)}e_ a\wedge d\delta\omega+\frac{p+1}{p}R_{ba}\wedge i_{X^b}\omega\nonumber\\
\nabla_{X_a}\delta\omega&=&-\frac{n-p+1}{(p+1)(n-p)}i_{X_a}\delta d\omega\nonumber\\
&&+\frac{n-p+1}{n-p}\left(i_{X_b}P_a\wedge i_{X^b}\omega+i_{X_b}R_{ca}\wedge i_{X^c}i_{X^b}\omega\right)\\
\frac{p}{p+1}\delta d\omega&+&\frac{n-p}{n-p+1}d\delta\omega=P_a\wedge i_{X^a}\omega+R_{ab}\wedge i_{X^a}i_{X^b}\omega.\nonumber
\end{eqnarray}
For a general CKY $p$-form, these integrability conditins are too restrictive for the equality of the coefficients of $\psi$ in (19) and (23). So, we will also consider a subset of CKY forms which are called normal CKY forms and defined by the following integrability conditions in terms of the 1-form $K_a$ defined in (18) \cite{Leitner, Ertem1}
\begin{eqnarray}
\nabla_{X_a}d\omega&=&\frac{p+1}{p(n-p+1)}e_a\wedge d\delta\omega+2(p+1)K_a\wedge\omega\nonumber\\
\nabla_{X_a}\delta\omega&=&-\frac{n-p+1}{(p+1)(n-p)}i_{X_a}\delta d\omega-2(n-p+1)i_{X^b}K_a\wedge i_{X_b}\omega\nonumber\\
\frac{p}{p+1}\delta d\omega&+&\frac{n-p}{n-p+1}d\delta\omega=-2(n-p)K_a\wedge i_{X^a}\omega.
\end{eqnarray}
In constant curvature backgrounds, the curvature 2-forms and Ricci 1-forms are written as $R_{ab}=\frac{\cal{R}}{n(n-1)}e_a\wedge e_b$ and $P_a=\frac{\cal{R}}{n}e_a$. One can also write the 1-form $K_a$ as $K_a=-\frac{\cal{R}}{2n(n-1)}e_a$. So, the integrability conditions (24) and (25) are equal to each other and all CKY forms are normal CKY forms in that case. Moreover, an Einstein manifold is defined in terms of Ricci 1-forms as $P_a=\frac{\cal{R}}{n}e_a$ and the integrability conditions of normal CKY forms in Einstein manifolds are equal to the integrability conditions of CKY forms in constant curvature backgrounds. Hence the integrability conditions of CKY forms in constant curvature backgrounds and of normal CKY forms in Einstein manifolds can be written as follows
\begin{eqnarray}
\nabla_{X_a}d\omega&=&\frac{p+1}{p(n-p+1)}e_a\wedge d\delta\omega-\frac{p+1}{n(n-1)}{\cal{R}}e_a\wedge\omega\nonumber\\
\nabla_{X_a}\delta\omega&=&-\frac{n-p+1}{(p+1)(n-p)}i_{X_a}\delta d\omega+\frac{n-p+1}{n(n-1)}{\cal{R}}i_{X_a}\omega\\
\frac{p}{p+1}\delta d\omega&+&\frac{n-p}{n-p+1}d\delta\omega=\frac{p(n-p)}{n(n-1)}{\cal{R}}\omega.\nonumber
\end{eqnarray}
We will show that the operator defined in (13) is a symmetry operator of the twistor equation for all CKY forms in constant curvature backgrounds and normal CKY forms in Einstein manifolds. So, the equality of the coefficients of $\psi$ in (19) and (23) is relevant in those cases. In the case of (19), the coefficient of $\psi$ is written as
\begin{eqnarray}
&&-(-1)^p\frac{p}{2}\omega.K_a+\frac{p}{2(p+1)}\nabla_{X_a}d\omega+\frac{p}{2(n-p+1)}\nabla_{X_a}\delta\omega\nonumber\\
&=&\frac{p}{4n(n-1)}{\cal{R}}\left(e_a\wedge\omega-i_{X_a}\omega\right)\nonumber\\
&&+\frac{1}{2(n-p+1)}e_a\wedge d\delta\omega-\frac{p}{2n(n-1)}{\cal{R}}e_a\wedge\omega\nonumber\\
&&-\frac{p}{2(p+1)(n-p)}i_{X_a}\delta d\omega+\frac{p}{2n(n-1)}{\cal{R}}i_{X_a}\omega\nonumber\\
&=&\frac{1}{2(n-p+1)}e_a\wedge d\delta\omega-\frac{p}{2(p+1)(n-p)}i_{X_a}\delta d\omega\nonumber\\
&&-\frac{p}{4n(n-1)}{\cal{R}}\left(e_a\wedge\omega-i_{X_a}\omega\right).
\end{eqnarray}
To obtain the coefficient of $\psi$ in (23), we expand the Clifford products in terms of wedge products. The first term is written as
\begin{eqnarray}
-(-1)^p\frac{p}{2n}e_a.e_b.\omega.K^b&=&-\frac{p}{2n}\bigg[(i_{X_b}K^b)e_a\wedge\omega+(i_{X_b}K^b)i_{X_a}\omega\nonumber\\
&&-e_a\wedge K^b\wedge i_{X_b}\omega-i_{X_a}(K^b\wedge i_{X_b}\omega)\nonumber\\
&&-(i_{X^c}K^b)e_a\wedge e_b\wedge i_{X_c}\omega-(i_{X^c}K^b)i_{X_a}(e_b\wedge i_{X_c}\omega)\nonumber\\
&&-(i_{X^c}K^b)e_a\wedge i_{X_b}i_{X_c}\omega-(i_{X^c}K^b)i_{X_a}i_{X_b}i_{X_c}\omega\bigg]\nonumber\\
&=&\frac{p(n-2p)}{4n^2(n-1)}{\cal{R}}\left(e_a\wedge\omega+i_{X_a}\omega\right).
\end{eqnarray}
For the other terms in (23), the following equalities can be written by using the definitions $e_a\wedge\nabla_{X^a}=d$ and $-i_{X_a}\nabla_{X^a}=\delta$ with the properties $d^2=0=\delta^2$ and the third equality in (26)
\begin{eqnarray}
e_a.e_b.\nabla_{X^b}d\omega&=&e_a.(i_{X_b}\nabla_{X^b}d\omega)\nonumber\\
&=&-e_a.\delta d\omega\nonumber\\
&=&-e_a\wedge\delta d\omega-i_{X_a}\delta d\omega\nonumber\\
&=&\frac{(p+1)(n-p)}{p(n-p+1)}e_a\wedge d\delta\omega\nonumber\\
&&-\frac{(p+1)(n-p)}{n(n-1)}{\cal{R}}e_a\wedge\omega-i_{X_a}\delta d\omega
\end{eqnarray}
and
\begin{eqnarray}
e_a.e_b.\nabla_{X^b}\delta\omega&=&e_a.(e_b\wedge\nabla_{X^b}\delta\omega)\nonumber\\
&=&e_a.d\delta\omega\nonumber\\
&=&e_a\wedge d\delta\omega+i_{X_a}d\delta\omega\nonumber\\
&=&e_a\wedge d\delta\omega-\frac{p(n-p+1)}{(p+1)(n-p)}i_{X_a}\delta d\omega\nonumber\\
&&+\frac{p(n-p+1)}{n(n-1)}{\cal{R}}i_{X_a}\omega.\nonumber\\
\end{eqnarray}
So, the coefficient of $\psi$ in (23) gives us
\begin{eqnarray}
&&-(-1)^p\frac{p}{2n}e_a.e^b.\omega.K_b+\frac{p}{2n(p+1)}e_a.e^b.\nabla_{X_b}d\omega\nonumber\\
&&+\frac{p}{2n(n-p+1)}e_a.e^b.\nabla_{X_b}\delta\omega\nonumber\\
&=&\frac{1}{2(n-p+1)}e_a\wedge d\delta\omega-\frac{p}{2(p+1)(n-p)}i_{X_a}\delta d\omega\nonumber\\
&&-\frac{p}{4n(n-1)}{\cal{R}}\left(e_a\wedge\omega-i_{X_a}\omega\right)
\end{eqnarray}
which is exactly equal to (27). This implies that the coefficients of $\psi$ in (19) and (23) are equal to each other for CKY forms in constant curvature backgrounds and normal CKY forms in Einstein manifolds. Thus, we have proven the equality of (19) and (23) and shown that the operator defined in (13) which is constructed from CKY forms is a symmetry operator of the twistor equation for CKY forms in constant curvature backgrounds and normal CKY forms in Einstein manifolds.

More generally, we can also prove that the operator defined in (13) is a symmetry operator of twistor spinors in all conformally-flat manifolds. Conformal 2-forms (or Weyl 2-forms) are defined for dimension $n>2$ manifolds in terms of the curvature characteristics as follows
\begin{equation}
C_{ab}=R_{ab}-\frac{1}{n-2}(P_a\wedge e_b-P_b\wedge e_a)+\frac{1}{(n-1)(n-2)}\mathcal{R}e_a\wedge e_b
\end{equation}
or if we write in terms of the 1-form $K_a$ defined in (18), we have
\begin{equation}
C_{ab}=R_{ab}+K_a\wedge e_b-K_b\wedge e_a.
\end{equation}
In conformally-flat manifolds, conformal 2-forms vanish $C_{ab}=0$ and we can write
\begin{equation}
R_{ab}=e_b\wedge K_a-e_a\wedge K_b.
\end{equation}
By contracting this with $X^a$, one finds
\begin{equation}
P_b=-(n-2)K_b+\frac{1}{2(n-1)}\mathcal{R}e_b.
\end{equation}
So, by using the above equalities, we can write the curvature terms in the integrability conditions of CKY equation given in (24) as follows
\begin{eqnarray}
\frac{p+1}{p}R_{ba}\wedge i_{X^b}\omega&=&(p+1)K_a\wedge\omega-\frac{p+1}{p}K_b\wedge e_a\wedge i_{X^b}\omega\nonumber\\
\frac{n-p+1}{n-p}i_{X_b}P_a\wedge i_{X^b}\omega&=&\frac{n-p+1}{n-p}\bigg(-(n-2)i_{X_b}K_a\wedge i_{X^b}\omega+\frac{1}{2(n-1)}\mathcal{R}i_{X_a}\omega\bigg)\nonumber\\
\frac{n-p+1}{n-p}i_{X_b}R_{ca}\wedge i_{X^c}i_{X^b}\omega&=&\frac{n-p+1}{n-p}\bigg(K_b\wedge i_{X^b}i_{X_a}\omega+(p-1)i_{X_b}K_a\wedge i_{X^b}\omega\bigg)\\
P_a\wedge i_{X^a}\omega&=&-(n-2)K_a\wedge i_{X^a}\omega+\frac{p}{2(n-1)}\mathcal{R}\omega\nonumber\\
R_{ab}\wedge i_{X^a}i_{X^b}\omega&=&2(p-1)K_a\wedge i_{X^a}\omega\nonumber
\end{eqnarray}
where we have used the equality $e_a\wedge i_{X_b}K_c\wedge i_{X^c}i_{X^b}\omega=0$ since $i_{X_b}K_c$ is symmetric in $b$ and $c$ while $i_{X^c}i_{X^b}$ is antisymmetric. Then, we can write the integrability conditions of the CKY equation in conformally-flat manifolds in the following form
\begin{eqnarray}
\nabla_{X_a}d\omega&=&\frac{p+1}{p(n-p+1)}e_a\wedge d\delta\omega+(p+1)K_a\wedge\omega-\frac{p+1}{p}K_b\wedge e_a\wedge i_{X^b}\omega\nonumber\\
\nabla_{X_a}\delta\omega&=&-\frac{n-p+1}{(p+1)(n-p)}i_{X_a}\delta d\omega+\frac{n-p+1}{n-p}\bigg(K_b\wedge i_{X^b}i_{X^a}\omega\nonumber\\
&&-(n-p-1)i_{X_b}K_a)\wedge i_{X^a}\omega+\frac{1}{2(n-1)}\mathcal{R}i_{X_a}\omega\bigg)\\
\frac{p}{p+1}\delta d\omega&+&\frac{n-p}{n-p+1}d\delta\omega=-(n-2p)K_a\wedge i_{X^a}\omega+\frac{p}{2(n-1)}\mathcal{R}\omega.\nonumber
\end{eqnarray}
Now, we can check the equivalence of (19) and (23) for the conformally-flat case. The coefficients of $\displaystyle{\not}D\psi$ are already equal to each other. So, we only need to check the equality of the coefficients of $\psi$. By using the equalities in (37), the coefficient of $\psi$ in (19) can be written as
\begin{eqnarray}
&&-(-1)^p\frac{p}{2}\omega.K_a+\frac{p}{2(p+1)}\nabla_{X_a}d\omega+\frac{p}{2(n-p+1)}\nabla_{X_a}\delta\omega\nonumber\\
&=&\frac{1}{2(n-p+1)}e_a\wedge d\delta\omega-\frac{p}{2(p+1)(n-p)}i_{X_a}\delta d\omega-\frac{1}{2}K_b\wedge e_a\wedge i_{X^b}\omega\nonumber\\
&&+\frac{p}{2(n-p)}\left((i_{X^b}K_a)i_{X_b}\omega+K_b\wedge i_{X^b}i_{X_a}\omega+\frac{1}{2(n-1)}\mathcal{R}i_{X_a}\omega\right)
\end{eqnarray}
and with the similar manipulations as in (28)-(30), the coefficient of $\psi$ in (23) can be found from (37) as
\begin{eqnarray}
&&-(-1)^p\frac{p}{2n}e_a.e_b.\omega.K^b+\frac{p}{2n(p+1)}e_a.e^b.\nabla_{X_b}d\omega+\frac{p}{2n(n-p+1)}e_a.e^b.\nabla_{X_b}\delta\omega\nonumber\\
&=&\frac{1}{2(n-p+1)}e_a\wedge d\delta\omega-\frac{p}{2(p+1)(n-p)}i_{X_a}\delta d\omega-\frac{1}{2}K_b\wedge e_a\wedge i_{X^b}\omega\nonumber\\
&&+\frac{p}{2(n-p)}\left((i_{X^b}K_a)i_{X_b}\omega+K_b\wedge i_{X^b}i_{X_a}\omega+\frac{1}{2(n-1)}\mathcal{R}i_{X_a}\omega\right).
\end{eqnarray}
So, from (38) and (39), the coefficients of $\psi$ in (19) and (23) are equal to each other and the operator defined in (13) is a symmetry operator of twistor spinors in conformally-flat backgrounds. For the special case of $n=3$, the conformal 2-forms $C_{ab}$ automatically vanishes in all manifolds. This means that the operator defined in (13) is a symmetry operator of twistor spinors for all manifolds in three dimensions.

As a result, we have proven the following
\begin{theorem}
Let $M$ be a $n$-dimensional spin manifold. For a CKY $p$-form $\omega$ on $M$ satisfying (8), the following first-order operator
\[
L_{\omega}=-(-1)^p\frac{p}{n}\omega.\displaystyle{\not}D+\frac{p}{2(p+1)}d\omega+\frac{p}{2(n-p+1)}\delta\omega
\]
is a symmetry operator for the twistor equation (4) in conformally-flat manifolds. Especially, it is constructed from all CKY forms in constant curvature manifolds and from normal CKY forms in Einstein manifolds.
\end{theorem}

Moreover, it is known that if a Riemannian spin manifold $M$ carries a twistor spinor with zero and its 1-form Dirac current corresponds to a non-trivial conformal Killing vector field, then the manifold $M$ is conformally-flat \cite{Kuhnel Rademacher}. So, the existence of a twistor spinor with zero whose Dirac current is a non-trivial conformal Killing vector field implies the symmetry operator property of (13).

In general, for a differential operator $O$ defining an equation, if its symmetry operator $L$ factors through $O$, namely $L=TO$ for some operator $T$, then $L$ is called a trivial symmetry of $O$. In our case, the symmetry operator in (13) cannot be written in terms of the twistor operator and so it is not a trivial operator.

\subsection{Conformal invariance}

Twistor operator defined in (3) is a conformally invariant operator under the rescalings of the metric and this fact leads to the conformal invariance of both CKY forms and the symmetry operators of twistor spinors given in (13). If we denote the conformal change of the metric g with the exponential function $e^{2\lambda}$ as $\widehat{g}=e^{2\lambda}g$, then the frame and co-frame basis and the connection 1-forms $\omega_{ab}$ are related by
\begin{eqnarray}
\widehat{X_a}=e^{-\lambda}X_a\quad\quad,\quad\quad \widehat{e^a}=e^{\lambda}e^a.\nonumber\\
\widehat{\omega}_{ab}=\omega_{ab}+X_b(\lambda)e_a-X_a(\lambda)e_b.\nonumber
\end{eqnarray}
Here, the exponential function and the co-frame basis can be distinguished by the usage of the function $\lambda$ and the co-frame indices. The relation between covariant derivatives with respect to a vector field $X$ acting on a $p$-form $\alpha$ in terms of $g$ and $\widehat{g}$ is given by
\begin{equation}
\widehat{\nabla}_X\alpha=\nabla_X\alpha-d\lambda\wedge i_X\alpha+\widetilde{X}\wedge i_{\widetilde{d\lambda}}\alpha.
\end{equation}
From this relation, one can also find the relations between exterior and co-derivatives
\begin{eqnarray}
\widehat{d}\alpha&=&d\alpha+pd\lambda\wedge\alpha\\
\widehat{\delta}\alpha&=&\delta\alpha-(n-p)i_{\widetilde{d\lambda}}\alpha.
\end{eqnarray}
Moreover, the covariant derivatives acting on a spinor field $\psi$ are related by
\begin{equation}
\widehat{\nabla}_{X_a}\psi=\nabla_{X_a}\psi+\frac{1}{2}e_a.d\lambda.\psi-\frac{1}{2}(i_{X_a}d\lambda).\psi
\end{equation}
and hence the relation between Dirac operators is
\begin{equation}
\widehat{\displaystyle{\not}D}\psi=\displaystyle{\not}D\psi+\frac{n-1}{2}d\lambda.\psi.
\end{equation}
From (43) and (44), one can easily see that the twistor equation (4) is conformally invariant and for a twistor spinor $\psi$ with respect to the metric $g$, the spinor $\widehat{\psi}=e^{\lambda/2}\psi$ is also a twistor spinor with respect to the rescaled metric $\widehat{g}$, namely we have the equality
\begin{equation}
\widehat{\nabla}_X\widehat{\psi}-\frac{1}{n}\widetilde{X}.\widehat{\displaystyle{\not}D}\widehat{\psi}=e^{\lambda/2}\left(\nabla_X\psi-\frac{1}{n}\widetilde{X}.\displaystyle{\not}D\psi\right).
\end{equation}
Similarly, from (40)-(42), the CKY equation (8) is also conformally invariant and if $\omega$ is a CKY $p$-form with respect to $g$, then $\widehat{\omega}=e^{(p+1)\lambda}\omega$ is also a CKY $p$-form with respect to $\widehat{g}$ and we have
\begin{eqnarray}
&&\widehat{\nabla}_X\widehat{\omega}-\frac{1}{p+1}i_X\widehat{d}\widehat{\omega}+\frac{1}{n-p+1}\widetilde{X}\wedge\widehat{\delta}\wedge\widehat{\omega}=e^{(p+1)\lambda}\left(\nabla_X\omega-\frac{1}{p+1}i_Xd\omega+\frac{1}{n-p+1}\widetilde{X}\wedge\delta\omega\right).
\end{eqnarray}
From these considerations we can say that the symmetry operators of twistor spinors given in (13) are constructed from conformally invariant objects and we can state the following
\begin{theorem}
For a twistor spinor $\psi$ on a $n$-dimensional spin manifold $M$, the symmetry operator $L_{\omega}$ defined in (13) in terms of CKY $p$-forms $\omega$ on $M$ is a conformally invariant operator.
\end{theorem}

\textbf{Proof}. For the relations (41), (42), (44), $\widehat{g}=e^{2\lambda}g$, $\widehat{\psi}=e^{\lambda/2}\psi$ and $\widehat{\omega}=e^{(p+1)\lambda}\omega$ as given above, we can calculate all the terms in the symmetry operator (13) written in the rescaled metric as
\begin{eqnarray}
-(-1)^p\frac{p}{n}\widehat{\omega}.\widehat{\displaystyle{\not}D}\widehat{\psi}&=&-e^{(p+3/2)\lambda}\frac{p}{n}\eta\omega.\displaystyle{\not}D\psi-\frac{p}{2}e^{(p+3/2)\lambda}\eta\omega.d\lambda.\psi\nonumber\\
\frac{p}{2(p+1)}\widehat{d}\widehat{\omega}.\widehat{\psi}&=&\frac{p}{2(p+1)}e^{(p+3/2)\lambda}d\omega.\psi+\frac{p}{2}e^{(p+3/2)\lambda}(d\lambda\wedge\omega).\psi\nonumber\\
\frac{p}{2(n-p+1)}\widehat{\delta}\widehat{\omega}.\widehat{\psi}&=&\frac{p}{2(n-p+1)}e^{(p+3/2)\lambda}\delta\omega.\psi-\frac{p}{2}e^{(p+3/2)\lambda}i_{\widetilde{d\lambda}}\omega.\psi.\nonumber
\end{eqnarray}
From the equality $\eta\omega.d\lambda=d\lambda\wedge\omega-i_{\widetilde{d\lambda}}\omega$, we can write
\begin{eqnarray}
&&-(-1)^p\frac{p}{n}\widehat{\omega}.\widehat{\displaystyle{\not}D}\widehat{\psi}+\frac{p}{2(p+1)}\widehat{d}\widehat{\omega}.\widehat{\psi}+\frac{p}{2(n-p+1)}\widehat{\delta}\widehat{\omega}.\widehat{\psi}\nonumber\\
&=&e^{(p+3/2)\lambda}\left(-(-1)^p\frac{p}{n}\omega.\displaystyle{\not}D\psi+\frac{p}{2(p+1)}d\omega.\psi+\frac{p}{2(n-p+1)}\delta\omega.\psi\right)\nonumber\\
\end{eqnarray}
and this proves that the symmetry operator in (13) is a conformally invariant operator.

\subsection{Commutator of symmetry operators}

Symmetry operators of an equation can provide a solution technique for that equation. If one finds a complete set of mutually commuting symmetry operators, then the equation can be solved by the method of separation of variables \cite{Miller}. So, we will investigate the commutators of symmetry operators of the twistor equation. By applying symmetry operators $L_{\omega_1}$ and $L_{\omega_2}$ which are constructed from a CKY $p$-form $\omega_1$ and a CKY $q$-form $\omega_2$ to a twistor spinor $\psi$, one obtains from (13)
\begin{eqnarray}
L_{\omega_1}L_{\omega_2}\psi&=&L_{\omega_1}\left(-(-1)^q\frac{q}{n}\omega_2.\displaystyle{\not}D\psi+\frac{q}{2(q+1)}d\omega_2.\psi+\frac{q}{2(n-q+1)}\delta\omega_2.\psi\right)\nonumber\\
&=&-(-1)^p\frac{p}{n}\omega_1.\displaystyle{\not}D\bigg(-(-1)^q\frac{q}{n}\omega_2.\displaystyle{\not}D\psi+\frac{q}{2(q+1)}d\omega_2.\psi+\frac{q}{2(n-q+1)}\delta\omega_2.\psi\bigg)\nonumber\\
&&+\frac{p}{2(p+1)}d\omega_1.\bigg(-(-1)^q\frac{q}{n}\omega_2.\displaystyle{\not}D\psi+\frac{q}{2(q+1)}d\omega_2.\psi+\frac{q}{2(n-q+1)}\delta\omega_2.\psi\bigg)\nonumber\\
&&+\frac{p}{2(n-p+1)}\delta\omega_1.\bigg(-(-1)^q\frac{q}{n}\omega_2.\displaystyle{\not}D\psi+\frac{q}{2(q+1)}d\omega_2.\psi+\frac{q}{2(n-q+1)}\delta\omega_2.\psi\bigg)\nonumber
\end{eqnarray}
which is equivalent to
\begin{eqnarray}
L_{\omega_1}L_{\omega_2}\psi&=&(-1)^{p+q}\frac{pq}{n^2}\omega_1.\displaystyle{\not}D\left(\omega_2.\displaystyle{\not}D\psi\right)-(-1)^p\frac{pq}{2n(q+1)}\omega_1.\displaystyle{\not}D\left(d\omega_2.\psi\right)\nonumber\\
&&-(-1)^p\frac{pq}{2n(n-q+1)}\omega_1.\displaystyle{\not}D\left(\delta\omega_2.\psi\right)-(-1)^q\frac{pq}{2n(p+1)}d\omega_1.\omega_2.\displaystyle{\not}D\psi\nonumber\\
&&+\frac{pq}{4(p+1)(q+1)}d\omega_1.d\omega_2.\psi+\frac{pq}{4(p+1)(n-q+1)}d\omega_1.\delta\omega_2.\psi\nonumber\\
&&-(-1)^q\frac{pq}{2n(n-p+1)}\delta\omega_1.\omega_2.\displaystyle{\not}D\psi+\frac{pq}{4(q+1)(n-p+1)}\delta\omega_1.d\omega_2.\psi\nonumber\\
&&+\frac{pq}{4(n-p+1)(n-q+1)}\delta\omega_1.\delta\omega_2.\psi.
\end{eqnarray}
Now, we compute the action of the Dirac operator on the Clifford product of a differential form and a spinor. So, we have the following equalities by using the integrability condition (17), definition of $K_a$ in (18) and the definition of the Hodge-de Rham operator on forms as $\displaystyle{\not}d=e^a.\nabla_{X_a}=d-\delta$
\begin{eqnarray}
\displaystyle{\not}D\left(\omega_2.\displaystyle{\not}D\psi\right)&=&e^a.\nabla_{X_a}\left(\omega_2.\displaystyle{\not}D\psi\right)\nonumber\\
&=&\displaystyle{\not}d\omega_2.\displaystyle{\not}D\psi+e^a.\omega_2.\nabla_{X_a}\displaystyle{\not}D\psi\nonumber\\
&=&d\omega_2.\displaystyle{\not}D\psi-\delta\omega_2.\displaystyle{\not}D\psi+e^a.\omega_2.\nabla_{X_a}\displaystyle{\not}D\psi\nonumber\\
&=&d\omega_2.\displaystyle{\not}D\psi-\delta\omega_2.\displaystyle{\not}D\psi-\frac{n}{2(n-2)}e^a.\omega_2.P_a.\psi\nonumber\\
&&+(-1)^q\frac{n(n-2q)}{4(n-1)(n-2)}{\cal{R}}\omega_2.\psi
\end{eqnarray}
where we have used $e^a.\omega_2.e_a=(-1)^q(n-2q)\omega_2$. Similarly, one can also find the equalities
\begin{eqnarray}
\displaystyle{\not}D\left(d\omega_2.\psi\right)&=&e^a.\nabla_{X_a}\left(d\omega_2.\psi\right)\nonumber\\
&=&\displaystyle{\not}d d\omega_2.\psi+e^a.d\omega_2.\nabla_{X_a}\psi\nonumber\\
&=&-\delta d\omega_2.\psi+\frac{1}{n}e^a.d\omega_2.e_a.\displaystyle{\not}D\psi\nonumber\\
&=&-\delta d\omega_2.\psi-\frac{(-1)^q}{n}(n-2(q+1))d\omega_2.\displaystyle{\not}D\psi
\end{eqnarray}
and
\begin{eqnarray}
\displaystyle{\not}D\left(\delta\omega_2.\psi\right)&=&e^a.\nabla_{X_a}\left(\delta\omega_2.\psi\right)\nonumber\\
&=&\displaystyle{\not}d\delta\omega_2.\psi+e^a.\delta\omega_2.\nabla_{X_a}\psi\nonumber\\
&=&d\delta\omega_2.\psi+\frac{1}{n}e^a.\delta\omega_2.e_a.\displaystyle{\not}D\psi\nonumber\\
&=&d\delta\omega_2.\psi-\frac{(-1)^q}{n}(n-2(q-1))\delta\omega_2.\displaystyle{\not}D\psi
\end{eqnarray}
So, by using (49), (50) and (51) in (48) we obtain
\begin{eqnarray}
L_{\omega_1}L_{\omega_2}\psi&=&\bigg[(-1)^{p+q}\frac{pq}{2n(q+1)}\omega_1.d\omega_2-(-1)^{p+q}\frac{pq}{2n(n-q+1)}\omega_1.\delta\omega_2\nonumber\\
&&-(-1)^q\frac{pq}{2n(p+1)}d\omega_1.\omega_2-(-1)^q\frac{pq}{2n(n-p+1)}\delta\omega_1.\omega_2\bigg].\displaystyle{\not}D\psi\nonumber\\
&+&\bigg[-(-1)^p\frac{p}{2(n-q+1)}\omega_1.d\delta\omega_2+\frac{pq}{4(p+1)(q+1)}d\omega_1.d\omega_2\nonumber\\
&&+\frac{pq}{4(p+1)(n-q+1)}d\omega_1.\delta\omega_2+\frac{pq}{4(q+1)(n-p+1)}\delta\omega_1.d\omega_2\nonumber\\
&&+\frac{pq}{4(n-p+1)(n-q+1)}\delta\omega_1.\delta\omega_2\nonumber\\
&&+(-1)^p\frac{pq}{2n^2(n-1)}\left(n-q-\frac{(n-2q)}{2(n-2)}\right){\cal{R}}\omega_1.\omega_2\bigg].\psi
\end{eqnarray}
where we have used $P_a=\frac{\cal{R}}{n}e_a$ and the third equality in (26). Similar calculations give the action of symmetry operators in reverse order as follows
\begin{eqnarray}
L_{\omega_2}L_{\omega_1}\psi&=&\bigg[(-1)^{p+q}\frac{pq}{2n(p+1)}\omega_2.d\omega_1-(-1)^{p+q}\frac{pq}{2n(n-p+1)}\omega_2.\delta\omega_1\nonumber\\
&&-(-1)^p\frac{pq}{2n(q+1)}d\omega_2.\omega_1-(-1)^p\frac{pq}{2n(n-q+1)}\delta\omega_2.\omega_1\bigg].\displaystyle{\not}D\psi\nonumber\\
&+&\bigg[-(-1)^q\frac{q}{2(n-p+1)}\omega_2.d\delta\omega_1+\frac{pq}{4(p+1)(q+1)}d\omega_2.d\omega_1\nonumber\\
&&+\frac{pq}{4(q+1)(n-p+1)}d\omega_2.\delta\omega_1+\frac{pq}{4(p+1)(n-q+1)}\delta\omega_2.d\omega_1\nonumber\\
&&+\frac{pq}{4(n-p+1)(n-q+1)}\delta\omega_2.\delta\omega_1\nonumber\\
&&+(-1)^q\frac{pq}{2n^2(n-1)}\left(n-p-\frac{(n-2p)}{2(n-2)}\right){\cal{R}}\omega_2.\omega_1\bigg].\psi
\end{eqnarray}
Hence, from (52) and (53), we can write the action of the commutator of two symmetry operators on a twistor spinor in the following form
\begin{eqnarray}
[L_{\omega_1}, L_{\omega_2}]\psi&=&\bigg[(-1)^{p+q}\frac{pq}{2n(q+1)}[\omega_1, d\omega_2]_{gCl}-(-1)^{p+q}\frac{pq}{2n(n-q+1)}[\omega_1, \delta\omega_2]_{+gCl}\nonumber\\
&&-(-1)^{p+q}\frac{pq}{2n(p+1)}[\omega_2, d\omega_1]_{gCl}+(-1)^{p+q}\frac{pq}{2n(n-p+1)}[\omega_2, \delta\omega_1]_{+gCl}\bigg].\displaystyle{\not}D\psi\nonumber\\
&+&\bigg[-(-1)^p\frac{p}{2(n-q+1)}\omega_1.d\delta\omega_2+(-1)^q\frac{q}{2(n-p+1)}\omega_2.d\delta\omega_1\nonumber\\
&&+\frac{pq}{4(p+1)(q+1)}[d\omega_1, d\omega_2]_{Cl}+\frac{pq}{4(p+1)(n-q+1)}[d\omega_1, \delta\omega_2]_{Cl}\nonumber\\
&&+\frac{pq}{4(q+1)(n-p+1)}[\delta\omega_1, d\omega_2]_{Cl}+\frac{pq}{4(n-p+1)(n-q+1)}[\delta\omega_1, \delta\omega_2]_{Cl}\nonumber\\
&&+\frac{pq}{2n^2(n-1)}{\cal{R}}\bigg((-1)^p\left(n-q-\frac{(n-2q)}{2(n-2)}\right)\omega_1.\omega_2\nonumber\\
&&-(-1)^q\left(n-p-\frac{(n-2p)}{2(n-2)}\right)\omega_2.\omega_1\bigg)\bigg].\psi
\end{eqnarray}
Here, we define the (graded) Clifford brackets for a $p$-form $\alpha$ and a $q$-form $\beta$ as
\begin{eqnarray}
[\alpha, \beta]_{Cl}&=&\alpha.\beta-\beta.\alpha\nonumber
\end{eqnarray}
\begin{eqnarray}
[\alpha, \beta]_{gCl}&=&\alpha.\beta-(-1)^q\beta.\alpha
\end{eqnarray}
\begin{eqnarray}
[\alpha, \beta]_{+gCl}&=&\alpha.\beta+(-1)^q\beta.\alpha\nonumber
\end{eqnarray}
Vanishing of the right hand side of (54) is a very restrictive condition on CKY forms in general. However, for the subsets of KY forms which satisfy $\delta\omega=0$ and closed CKY forms which satisfy $d\omega=0$, it can be written in a more simple form. For KY forms we have
\begin{eqnarray}
[L_{\omega_1}, L_{\omega_2}]\psi&=&(-1)^{p+q}\frac{pq}{2n}\bigg[\frac{1}{q+1}[\omega_1, d\omega_2]_{gCl}-\frac{1}{p+1}[\omega_2, d\omega_1]_{gCl}\bigg].\displaystyle{\not}D\psi\nonumber\\
&&+pq\bigg[\frac{1}{4(p+1)(q+1)}[d\omega_1, d\omega_2]_{Cl}+\frac{\cal{R}}{2n^2(n-1)}\bigg((-1)^p(n-q-\frac{(n-2q)}{2(n-2)})\omega_1.\omega_2\nonumber\\
&&-(-1)^q(n-p-\frac{(n-2p)}{2(n-2)})\omega_2.\omega_1\bigg)\bigg].\psi
\end{eqnarray}
For closed CKY forms, (54) reduces to
\begin{eqnarray}
[L_{\omega_1}, L_{\omega_2}]\psi&=&(-1)^{p+q}\frac{pq}{2n}\bigg[\frac{1}{n-p+1}[\omega_2, \delta\omega_1]_{+gCl}-\frac{1}{n-q+1}[\omega_1, \delta\omega_2]_{+gCl}\bigg].\displaystyle{\not}D\psi\nonumber\\
&&+pq\bigg[\frac{1}{4(n-p+1)(n-q+1)}[\delta\omega_1, \delta\omega_2]_{Cl}\nonumber\\
&&+\frac{\cal{R}}{2n^2(n-1)}\bigg(-(-1)^p\left(q+\frac{n}{q}+\frac{(n-2q)}{2(n-2)}\right)\omega_1.\omega_2\nonumber\\
&&+(-1)^q\left(p+\frac{n}{p}+\frac{(n-2p)}{2(n-2)}\right)\omega_2.\omega_1\bigg)\bigg].\psi
\end{eqnarray}
Moreover, for the flat Minkowski backgrounds, the curvature scalar vanishes ${\cal{R}}=0$. So, from the KY forms and closed CKY forms that satisfy the vanishing conditions of the non-curvature parts of (56) and (57) in Minkowski spacetime, one can find the mutually commuting symmetry operators of the twistor equation to solve it by the method of separation of variables. For KY forms, the conditions turn out to be
\begin{equation}
\frac{1}{q+1}[\omega_1, d\omega_2]_{gCl}=\frac{1}{p+1}[\omega_2, d\omega_1]_{gCl}
\end{equation}
\begin{equation}
[d\omega_1, d\omega_2]_{Cl}=0
\end{equation}
and for closed CKY forms, they are are equal to
\begin{equation}
\frac{1}{n-p+1}[\omega_2, \delta\omega_1]_{+gCl}=\frac{1}{n-q+1}[\omega_1, \delta\omega_2]_{+gCl}
\end{equation}
\begin{equation}
[\delta\omega_1, \delta\omega_2]_{Cl}=0.
\end{equation}
Because of the conformal invariance of the twistor equation, from the solutions of the twistor equation in Minkowski background, one can obtain the twistor spinors on maximally symmetric spacetimes of constant curvature since they correspond to conformally flat manifolds.

On the other hand, the solutions of the twistor equation can also be found from a known solution of the equation. Since the squaring map of twistor spinors give CKY forms, the symmetry operator defined in (13) can be written in terms of twistor spinors only. For a spinor $\psi$ and its dual $\overline{\psi}$ with respect to the inner product $(.,.)$ on spinors, the squaring map is defined as
\begin{eqnarray}
\psi\otimes\overline{\psi}&=&(\psi,\psi)+(\psi, e_a.\psi)e^a+(\psi, e_{ba}.\psi)e^{ab}+...+(\psi, e_{a_p...a_2a_1}.\psi)e^{a_1a_2...a_p}\nonumber\\
&&+...+(-1)^{\lfloor{n/2}\rfloor}(\psi, z.\psi)z
\end{eqnarray}
where $e^{a_1a_2...a_p}=e^{a_1}\wedge e^{a_2}\wedge...\wedge e^{a_p}$, $\lfloor{ }\rfloor$ is the floor function that takes the integer part of the argument and $z$ is the volume form. This equality is called the Fierz identity and gives the expression of spinor bilinears in terms of differential forms. The homogeneous $p$-form parts of the right hand side of (62) are called $p$-form Dirac currents and denoted by
\begin{equation}
(\psi\overline{\psi})_p=(\psi, e_{a_p...a_2a_1}.\psi)e^{a_1a_2...a_p}.
\end{equation}
For a twistor spinor $\psi$, the $p$-form Dirac currents correspond to CKY forms \cite{Acik Ertem}. This means that the spinor bilinear $\omega=\psi\otimes\overline{\psi}$ of a twistor spinor $\psi$ is a sum of CKY forms. So, the existence of twistor spinors implies the existence of CKY forms, namely their integrability conditions are related to each other. Then, the symmetry operators of the twistor equation can be written in terms of a twistor spinor $\psi$ as follows
\begin{equation}
L_{\psi\overline{\psi}}\psi=-(-1)^p\frac{p}{n}(\psi\overline{\psi})_p.\displaystyle{\not}D\psi+\frac{p}{2(p+1)}d(\psi\overline{\psi})_p.\psi+\frac{p}{2(n-p+1)}\delta(\psi\overline{\psi})_p.\psi
\end{equation}
Moreover, by using the twistor equation, the exterior derivative and co-derivative of the $p$-form Dirac currents of twistor spinors can be calculated and found as
\begin{equation}
d(\psi\overline{\psi})_p=\frac{p+1}{n}\bigg(\displaystyle{\not}d(\psi\overline{\psi})-2i_{X^a}(\psi\overline{\nabla_{X_a}\psi})\bigg)_{p+1}
\end{equation}
and
\begin{equation}
\delta(\psi\overline{\psi})_p=-\frac{n-p+1}{n}\bigg(\displaystyle{\not}d(\psi\overline{\psi})-2e^a\wedge(\psi\overline{\nabla_{X_a}\psi})\bigg)_{p-1}
\end{equation}
(see \cite{Acik Ertem} for the proofs). From these equalities, the symmetry operator in (64) can also be written in the following alternative form
\begin{eqnarray}
L_{\psi\overline{\psi}}\psi&=&-(-1)^p\frac{p}{n}(\psi\overline{\psi})_p.\displaystyle{\not}D\psi+\frac{p}{2n}\big(\displaystyle{\not}d(\psi\overline{\psi})\big)_{p+1}.\psi-\frac{p}{n}\big(i_{X^a}(\psi\overline{\nabla_{X_a}\psi})\big)_{p+1}.\psi\nonumber\\
&&-\frac{p}{2n}\big(\displaystyle{\not}d(\psi\overline{\psi})\big)_{p-1}.\psi+\frac{p}{n}\big(e^a\wedge(\psi\overline{\nabla_{X_a}\psi})\big)_{p-1}.\psi\nonumber\\
&=&-\frac{p}{n}\bigg[(-1)^p(\psi\overline{\psi})_p.\displaystyle{\not}D\psi+\frac{1}{n}\bigg(\big(i_{X^a}(\psi\overline{e_a.\displaystyle{\not}D\psi})\big)_{p+1}-\big(e^a\wedge(\psi\overline{e_a.\displaystyle{\not}D\psi})\big)_{p-1}\bigg).\psi\bigg]\nonumber\\
&&+\frac{p}{2n}\bigg[\big(\displaystyle{\not}d(\psi\overline{\psi})\big)_{p+1}-\big(\displaystyle{\not}d(\psi\overline{\psi})\big)_{p-1}\bigg].\psi
\end{eqnarray}
This gives an alternative method to obtain twistor spinors from a known solution without using CKY forms. In this case, we use only one twistor spinor and for $p=0$, $(\psi\overline{\psi})_p$ gives a function. So, the operator (67) may be a constant multiple of $\psi$ as a special case. For $p\neq 0$, $(\psi\overline{\psi})_p$ is a differential form and its action on a spinor gives another spinor in general. So, the operator (67) can be a first-order operator in this case. However, the strict form of (67) as a constant multiple or a first-order operator should be studied in a future research.

\section{Extended Conformal Superalgebras}

Twistor spinors and conformal Killing vectors which correspond to the metric duals of the 1-form Dirac currents of twistor spinors can be combined into a conformal superalgebra structure \cite{Habermann2}. A superalgebra $\mathfrak{g}=\mathfrak{g}_0\oplus\mathfrak{g}_1$ consists of an even part $\mathfrak{g}_0$ and the odd part $\mathfrak{g}_1$ with a bilinear multiplication operation
\begin{equation}
[.,.]:\mathfrak{g}_i\times\mathfrak{g}_j\longrightarrow\mathfrak{g}_{i+j}
\end{equation}
where $i,j=0,1$ mod 2. For the elements $a,b\in\mathfrak{g}$ the bilinear operation satisfies the property $[a,b]=-(-1)^{|a||b|}[b,a]$ where $|a|$ corresponds to 0 or 1 for $a$ is in $\mathfrak{g}_0$ or $\mathfrak{g}_1$. The even part of the conformal superalgebra corresponds to the Lie algebra of conformal Killing vectors and the odd part consists of the twistor spinors in the background. The brackets that correspond to the bilinear operation are defined as follows. The even-even bracket is the ordinary Lie bracket of vector fields
\begin{equation}
[.,.]:\mathfrak{g}_0\times\mathfrak{g}_0\longrightarrow\mathfrak{g}_0
\end{equation}
The even-odd bracket is the symmetry operator of twistor spinors which is defined from the Lie derivative of twistor spinors with respect to a conformal Killing vector in (11)
\begin{equation}
{\cal{L}}-\frac{1}{2}\lambda:\mathfrak{g}_0\times\mathfrak{g}_1\longrightarrow\mathfrak{g}_1
\end{equation}
The odd-odd bracket corresponds to the 1-form Dirac current of a twistor spinor, namely the 1-form part of the squaring map defined in (62)
\begin{equation}
(\,\,)_1:\mathfrak{g}_1\times\mathfrak{g}_1\longrightarrow\mathfrak{g}_0
\end{equation}
Moreover, this superalgebra can be extended to a Lie superalgebra by defining some extra R-symmetries in the background \cite{de Medeiros Hollands}. In a Lie superalgebra, the bilinear multiplication defined in (68) satisfies the graded Jacobi identities.

Obtaining the more general symmetry operators of twistor spinors gives way to the construction of extended conformal superalgebras in conformally flat backgrounds. Those extended superalgebras will include all of CKY forms besides conformal Killing vectors. For that aim, we need to define a bracket for CKY forms. Indeed, CKY forms satisfy a graded Lie algebra in constant curvature backgrounds and Einstein manifolds with respect to the following bracket defined for a CKY $p$-form $\omega_1$ and a CKY $q$-form $\omega_2$
\begin{eqnarray}
[\omega_1, \omega_2]_{CKY}&=&\frac{1}{q+1}i_{X_a}\omega_1\wedge i_{X^a}d\omega_2+\frac{(-1)^p}{p+1}i_{X_a}d\omega_1\wedge i_{X^a}\omega_2\nonumber\\
&&+\frac{(-1)^p}{n-q+1}\omega_1\wedge\delta\omega_2+\frac{1}{n-p+1}\delta\omega_1\wedge\omega_2
\end{eqnarray}
which is proved in \cite{Ertem1}. This bracket gives a CKY ($p+q-1$)-form. So, we can define an extended superalgebra $\bar{\mathfrak{g}}=\bar{\mathfrak{g}}_0\oplus\bar{\mathfrak{g}}_1$ and the even part corresponds to the graded Lie algebra of CKY forms and the odd part is the space of twistor spinors. Then, the even-even bracket is defined as
\begin{equation}
[.,.]_{CKY}:\bar{\mathfrak{g}}_0\times\bar{\mathfrak{g}}_0\longrightarrow\bar{\mathfrak{g}}_0
\end{equation}
We have proven in Section II that the operator defined in (13) is a symmetry operator for twistor spinors in conformally flat backgrounds. So, the even-odd bracket of the extended conformal superalgebra corresponds to the symmetry operator of twistor spinors
\begin{equation}
L:\bar{\mathfrak{g}}_0\times\bar{\mathfrak{g}}_1\longrightarrow\bar{\mathfrak{g}}_1
\end{equation}
For the odd-odd bracket, we have the $p$-form Dirac currents of twistor spinors defined in (63)
\begin{equation}
(\,)_p:\bar{\mathfrak{g}}_1\times\bar{\mathfrak{g}}_1\longrightarrow\bar{\mathfrak{g}}_0
\end{equation}
where $(\,\,)_p$ is a specific projection on each $p$-form. Since CKY bracket is a graded Lie bracket and $L$ is a symmetry operator for twistor spinors in conformally-flat manifolds, the extended conformal superalgebras can be defined in those cases by the brackets (73)-(75) and this is true for all manifolds in three dimensions. However, this construction does not correspond to a Lie superalgebra since the defined brackets do not satisfy the graded Jacobi identities. These extended conformal superalgebras may be the signs of new supersymmetric field theories in conformal backgrounds as in the case of conformal superalgebras.

Dimension of a superalgebra is denoted by $(\alpha|\beta)$ where $\alpha$ denotes the dimension of the even part and $\beta$ is the dimension of the odd part. CKY forms and twistor spinors have maximal number of dimensions in conformally flat backgrounds. This maximum number for CKY $p$-forms in $n$ dimensions is given by \cite{Semmelmann}
\begin{equation}
C_p=\left(
             \begin{array}{c}
               n \\
               p-1 \\
             \end{array}
           \right)+2\left(
                      \begin{array}{c}
                        n \\
                        p \\
                      \end{array}
                    \right)+\left(
                              \begin{array}{c}
                                n \\
                                p+1 \\
                              \end{array}
                            \right)
\end{equation}
and for twistor spinors, the maximum dimension is equal to
\begin{equation}
t=2^{\lfloor n/2\rfloor}+1.
\end{equation}
So, while the dimension of the conformal superalgebras is equal to $(C_1|t)$, the dimension for the extended conformal superalgebras is $(\sum_pC_p|t)$ in conformally flat backgrounds.

Killing vector fields constitute a Lie subalgebra in the Lie algebra of conformal Killing vectors and Killing spinors are twistor spinors which are eigenspinors of the Dirac operator at the same time. So, for the subsets of Killing vectors and Killing spinors, conformal superalgebras reduces to Killing superalgebras. Similarly, KY forms are in a subset of the space of CKY forms and are defined as co-closed CKY forms, namely they satisfy the condition $\delta\omega=0$. They constitute a graded Lie algebra structure with respect to Schouten-Nijenhuis (SN) bracket in constant curvature backgrounds \cite{Kastor Ray Traschen}. Extended conformal superalgebras defined for CKY forms and twistor spinors reduce to extended Killing superalgebras in which the even part corresponds to the Lie algebra of special KY forms and special closed CKY forms and odd part is the space of Killing spinors \cite{Ertem2, Acik Ertem2}. Indeed, the CKY bracket in (72) reduces to SN bracket for KY forms $\omega_1$ and $\omega_2$
\begin{equation}
[\omega_1,\omega_2]_{SN}=i_{X^a}\omega_1\wedge\nabla_{X_a}\omega_2+(-1)^pi_{X^a}\omega_2\wedge\nabla_{X_a}\omega_1
\end{equation}
and the symmetry operator for twistor spinors in (13) reduces to the symmetry operator of Killing spinors which is defined as
\begin{equation}
L_{\omega}=-(-1)^p\lambda p\omega.\psi+\frac{p}{2(p+1)}d\omega.\psi
\end{equation}
for a KY $p$-form $\omega$ \cite{Ertem2}. The Killing number $\lambda$ is a real or pure imaginary constant that comes from the Killing spinor equation for $\psi$
\begin{equation}
\nabla_X\psi=\lambda\widetilde{X}.\psi.
\end{equation}
The $p$-form Dirac currents of Killing spinors correspond to KY forms or their Hodge duals \cite{Acik Ertem, Acik Ertem2}. So, the extended conformal superalgebras constructed by (73)-(75) are generalizations of extended Killing superalgebras to CKY forms and twistor spinors.

\section{Conclusion}

The graded Lie algebra of CKY forms in constant curvature backgrounds and of normal CKY forms in Einstein manifolds can be constructed from a modified SN bracket \cite{Ertem1}. It is shown in this paper that these CKY forms can also be used in the construction of the symmetry operators of twistor spinors. Moreover, one can obtain the conditions to find mutually commuting symmetry operators of twistor spinors and compute the solutions of the twistor equation in conformally flat backgrounds. By using these constructions, conformal superalgebras of conformal Killing vector fields and twistor spinors are extended to include CKY forms. The brackets of extended conformal superalgebras correspond to the graded Lie bracket of CKY forms, symmetry operators of twistor spinors and the $p$-form Dirac currents of twistor spinors which are equal to CKY forms. They reduce to the extended Killing superalgebras for the cases of KY forms and Killing spinors.

Supersymmetric field theories in flat spacetimes can be extended to curved backgrounds by finding the conditions on the preserved supersymmetry parameters. For the case of supersymmetric field theories in conformal supergravity backgrounds, the supersymmetry parameters satisfy the twistor equation. This gives way to define conformal superalgebras for those supersymmetric theories in curved backgrounds. The construction of the extended conformal superalgebras can be used for finding supersymmetric field theories in conformally flat backgrounds related to these extended superalgebras.

The next step may be to consider supersymmetric gauge theories in curved backgrounds. In those cases, the supersymmetry parameters satisfy a gauged twistor equation in which the connection is modified by an extra gauge term. New extended superalgebras can also be investigated by considering the recently constructed symmetry operators of gauged twistor spinors in \cite{Ertem3}.

\begin{acknowledgments}

The author thanks Jose M. Figueroa-O`Farrill, Andrea Santi and \"{O}zg\"{u}r A\c{c}{\i}k for inspiring discussions on various subjects. He also thanks School of Mathematics of The University of Edinburgh for the kind hospitality and providing a fruitful scientific atmosphere during his stay in Edinburgh where this work started.

\end{acknowledgments}


\end{document}